\documentclass[twocolumn,preprintnumbers,prl,aps,amssymb,superscriptaddress]{revtex4}

\usepackage{graphicx}
\usepackage{dcolumn}
\usepackage{bm}

\begin{document}

\title{Electric Field Tuned Dimensional Crossover in Ar-Irradiated SrTiO$_3$}

\author{J.H. Ngai}
 \affiliation{Department of Applied Physics, and Center for Research on Interface Structures and Phenomena, Yale University, 15 Prospect Street, New Haven, CT  06520-8284, USA}
\author{Y. Segal}
 \affiliation{Department of Applied Physics, and Center for Research on Interface Structures and Phenomena, Yale University, 15 Prospect Street, New Haven, CT  06520-8284, USA}
\author{F.J. Walker}
 \affiliation{Department of Applied Physics, and Center for Research on Interface Structures and Phenomena, Yale University, 15 Prospect Street, New Haven, CT  06520-8284, USA}
\author{S. Ismail-Beigi}
 \affiliation{Department of Applied Physics, and Center for Research on Interface Structures and Phenomena, Yale University, 15 Prospect Street, New Haven, CT  06520-8284, USA}
 \affiliation{Department of Physics, Yale University, 217 Prospect Street, New Haven, CT  06511-8499, USA}
\author{K. Le Hur}
  \affiliation{Department of Applied Physics, and Center for Research on Interface Structures and Phenomena, Yale University, 15 Prospect Street, New Haven, CT  06520-8284, USA}
   \affiliation{Department of Physics, Yale University, 217 Prospect Street, New Haven, CT  06511-8499, USA}
\author{C. H. Ahn}%
 \affiliation{Department of Applied Physics, and Center for Research on Interface Structures and Phenomena, Yale University, 15 Prospect Street, New Haven, CT  06520-8284, USA}
 \affiliation{Department of Physics, Yale University, 217 Prospect Street, New Haven, CT  06511-8499, USA}

\date{\today}

\begin{abstract}

We present low temperature magnetoresistance measurements of Ar-irradiated SrTiO$_3$ under an applied electrostatic field. The electric field, applied through a back gate bias, modulates both the mobility and sheet density, with a greater effect on the former. For high mobilities, 3-dimensional orbital magnetoresistance is observed. For low mobilities, negative magnetoresistance that is consistent with the suppression of 2-dimensional weak-localization is observed. The crossover from 3 to 2-dimensional transport arises from a modulation in the carrier confinement, which is enhanced by the electric field dependent dielectric constant of SrTiO$_3$. The implications of our results on the development of oxide electronic devices will be discussed.

\end{abstract}

\maketitle
 
Transition metal oxides exhibit extraordinary functionality that promise to enhance the performance of future electronic devices. The various functionalities of these compounds can be tuned through charge carrier doping \cite{Ahn03}. A well known example of such a compound is SrTiO$_3$ (STO), which transitions from insulating to semi-conducting behavior as electrons are progressively doped into the lattice through oxygen vacancies or chemical substitution \cite{Tufte}. Recent advances have enabled doping to be locally confined near interfaces and surfaces, which has led to the study of quasi-2 dimensional electron gases (Q2DEG) in STO. In particular, much attention has focused on the Q2DEG found at the LaAlO$_3$ (LAO)/STO interface \cite{Hwang,Willmott}, where magnetic and superconducting ground states can be achieved, depending on growth conditions \cite{Hilgenkamp,Triscone08,Triscone09}. A Q2DEG can also be realized by Ar-irradiation, which dopes through the creation of oxygen vacancies near the irradiated surface \cite{Reagor,Kan,Klein,Butko}. 

The rich phenomena observed in both systems are intimately connected to the spatial confinement of the carriers, which affects the dimensionality of and scattering within the Q2DEG \cite{Siemons,Copie}. Tuning the confinement thus modulates the properties of a Q2DEG. At present, the confinement is largely determined by the initial growth or fabrication conditions of the Q2DEG \cite{Willmott}. Application of a back gate bias could enable the confinement to be tuned continuously, as demonstrated for 2DEG's in Si and GaAs \cite{Fowler,Hirakawa}. For the LAO/STO interface, Caviglia \emph{et al.} have successfully tuned the superconducting transition temperature $T_c$ through a back gate bias \cite{Triscone08}. The electrostatic tuning of $T_c$ has been attributed to changes in sheet carrier density $n_s$, induced by the large dielectric constant $\epsilon$ of STO \cite{Christen}. However, Bell \emph{et al.} \cite{Hwang09} suggest a modulation in confinement also contributes to the change in $T_c$. In this regard, elucidating the effect of a back gate bias on the confinement and properties of a Q2DEG in general, is important. Furthermore, combining the multiple functionalities of oxides could enable intrinsic properties, such as the dielectric constant of STO, to be utilized in unconventional ways.
 
In this Letter we present low temperature magnetoresistance (MR) measurements of Ar-irradiated STO under an applied electrostatic field. The electric field, created through a back gate bias $V_G$, modulates the mobility $\mu$ and $n_s$ of the Q2DEG, with a greater effect on the former. In the high $\mu$ regime, conventional MR that is 3D in nature is observed. For the low $\mu$ regime, negative MR that is consistent with the suppression of 2D weak-localization is observed. The evolution of the MR with $V_G$ arises from a modulation in the carrier confinement, which is enhanced by $\epsilon (E)$ of STO, as shown through numerical solutions to the Poisson and Schr\"odinger equations.
By combining the multiple functionalities of STO, we demonstrate that a magnetic field sensitive, mobility modulated device can be created.

\begin{figure}[t]
\centering
\includegraphics[width=8.5cm] {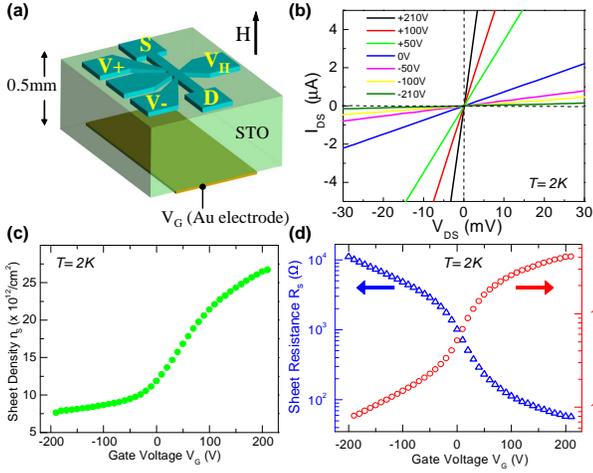}
\caption{\label{10509_Fig1} (color online) Electrical characterization of Ar-irradiated SrTiO$_3$ devices as a function of $V_G$ at $T$=2K. (a) Schematic of device. (b) Modulation of current-voltage characteristics with $V_G$. (c) Sheet carrier density versus $V_G$ showing a $\sim$3.5 factor change for -190V$\leq V_G\leq$210V.(d) $R_s$ ($\mu$) versus $V_G$ shown as triangles (circles). }
\end{figure}

A schematic of our Ar-irradiated devices is shown in Fig.\ref{10509_Fig1}(a). The 80 x 40$\mu$m Hall bar is defined using standard photolithography techniques on a STO substrate. The device becomes conducting after irradiation with 1keV Ar-ions at a rate of $\sim$3.7$\times$10$^{12}$ cm$^{-2}$s$^{-1}$ for 90 minutes. Ohmic contacts are formed by wirebonding Al wires directly to the STO, and the gate contact is made by depositing Au on the back side of the device. Sheet resistances $R_s$ are $\sim$9k$\Omega$ at 300K, and the devices exhibit metallic behavior upon cooling \cite{Butko}. At 2K, $V_G$ is first ramped between $\pm$210V to reduce charge trapping in the STO and minimize hysteresis in the gating. Gate leakage currents are $<$1nA for all $V_G$. Representative data from one of the devices are shown in the figures below. 

Figure \ref{10509_Fig1}(b) shows modulation of the 2-point current-voltage ($I_{DS}$-$V_{DS}$) characteristics with $V_G$. Figure \ref{10509_Fig1}(d) (triangles) indicates that $R_s$ can be modulated by over two orders of magnitude between -190V$\leq V_G\leq$210V. For positive (negative) $V_G$, $R_s$ decreases (increases), consistent with the addition (removal) of n-type carriers to (from) the channel of the device according to the relation $R_s= (n_se\mu)^{-1}$, where $e$ is the electron charge. Hall measurements at each $V_G$ directly determined $\Delta n_s$, which is related to the Hall coefficient through $\Delta R_H$=$(-\Delta n_se)^{-1}$. We find that $n_s$ can be tuned from $\sim$8 to $\sim$27x10$^{12}$cm$^{-2}$ for -190V$\leq V_G\leq$210V as shown in Fig.\ref{10509_Fig1}(c). This modest $\Delta n_s$ does not account for $\Delta R_s$, thus $V_G$ also modulates $\mu$, as shown in Fig.\ref{10509_Fig1}(d) (circles). For the device shown, $\mu$ can be modulated by a factor of $\sim$50, from $\sim$80 to $\sim$4100cm$^2$/Vs \cite{Ionic_conductivity}.  
 
\begin{figure}[t]
\centering
\includegraphics[width=7.5cm] {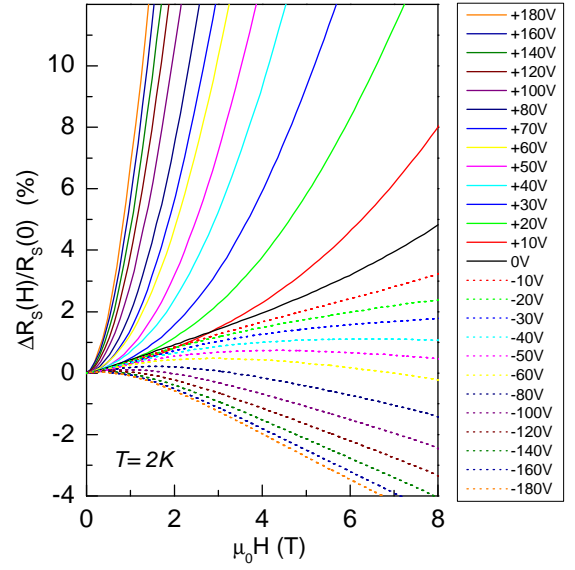}
\caption{\label{10509_Fig2} (color online) Electrostatic modulation of the MR of Ar-irradiated SrTiO$_3$ at $T$=2K. MR for positive (negative) $V_G$ are shown as solid (dotted) lines. A crossover from positive to negative MR is observed going from high to low $\mu$ regimes.}
\end{figure}

MR measurements provide insight on the modulation in $\mu$. The MR, defined as $\Delta R_s(H)$/$R_s(0)$ where $\Delta R_s(H)$=$R_s(H)$-$R_s(0)$, is shown in Fig.\ref{10509_Fig2} as a function of magnetic field $B$=$\mu_0 H$ applied out of the plane of the device for various $V_G$. Positive MR is observed (solid curves) in the high $\mu$ regime associated with $V_G>$0. In contrast, negative MR is observed (dotted curves) in the low $\mu$ regime associated with $V_G<<$0.

\begin{figure}[t]
\centering
\includegraphics[width=8.5cm] {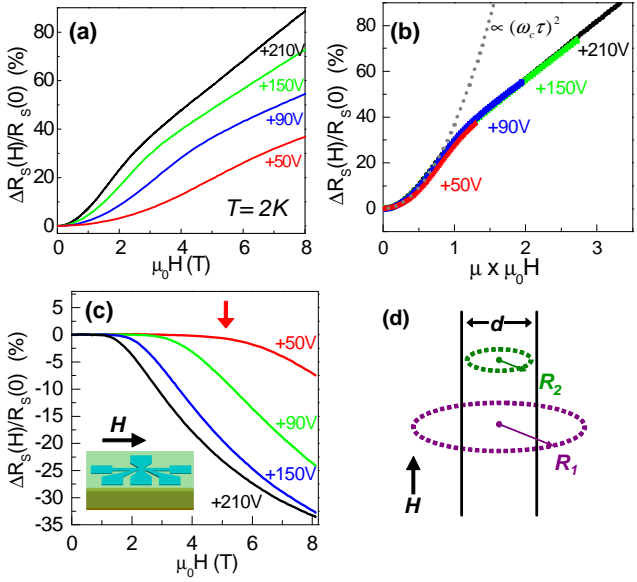}
\caption{\label{10509_Fig3} (color online) High $\mu$ regime MR of Ar-irradiated SrTiO$_3$ at $T$=2K. (a) Evolution of MR with $V_G$. (b) Kohler plot of MR with weak field $\propto(\omega_c\tau)^2$ fit (grey dotted). (c) In-plane MR. Red arrow indicates onset of negative MR for $V_G$= +50V. (d) Cross-sectional illustration of a thin conducting plate of thickness $d$. Orbital radius $R_1>(d/2)$ ($R_2<(d/2)$). }
\end{figure}

We analyze the high $\mu$ MR shown in Fig.\ref{10509_Fig3}(a) to elucidate the dimensionality of the electron gas in this regime. Figure \ref{10509_Fig3}(b) shows the MR data plotted against $\mu(V_G)\times B$. The data at each $V_G$ collapse virtually onto a common curve, indicating that Kohler's rule is obeyed, $i.e.$$\Delta R_s(B)/R_s(0)$=$F(B\tau)$, where $\tau$ is the carrier relaxation time. At weak magnetic fields, the MR is described by $F$=$A(\mu B)^2$= $A(\omega_c\tau)^2$ where $\omega_c$ is the cyclotron frequency as shown by the fit in Fig.\ref{10509_Fig3}(b) (grey dotted). We find that $A$=0.37$\pm$0.05, which agrees well with the value of $\sim$0.38 expected for MR that arises from a 3D electron gas \cite{Putley}.

The depth of the carrier confinement in the high $\mu$ regime can be estimated by examining the MR for magnetic fields applied parallel to the current in the plane of the device, as shown in the inset of Fig.\ref{10509_Fig3}(c). In this geometry, the Lorentz force associated with $B$ does not scatter the component of carrier momentum parallel to $I_{DS}$. For our devices, we observe negative MR that develops beyond a threshold field $B^*$, which varies with $V_G$. For example, $B^*\sim$5.5T for the $V_G$=+50V curve (red), as shown by the red arrow in Fig.\ref{10509_Fig3}(c). In high $\mu$, non-magnetic systems, negative in-plane MR can arise from a reduction in the scattering of charge carriers at the sample boundaries \cite{MacDonald}. In this context, the dependence of $B^*$ on $V_G$ is understood if we consider the channel as a plate of thickness $d$, as shown in Fig.\ref{10509_Fig3}(d). The charge carriers follow helical paths of radii $R\propto B^{-1}$ induced by the Lorentz force of the in-plane field. A reduction in boundary scattering occurs for carriers of orbital radii $R\leq d/2$. Thus, $d$ in the high $\mu$ regime can be estimated by relating $B^*$ to the classical cyclotron radius \cite{DepthEquation}  

\begin{equation}\label{Depth}d\sim2\left(\frac{3\pi^2n_s\hbar^3}{2e^3(B^*)^3}\right)^{1/4}.
\end{equation} 
From Eqn.\ref{Depth} we find that $d$ increases from $\sim$90$nm$ at $V_G$=+50V to $\sim$320$nm$ at $V_G$=+210V.

\begin{figure}[t]
\centering
\includegraphics[width=8.5cm] {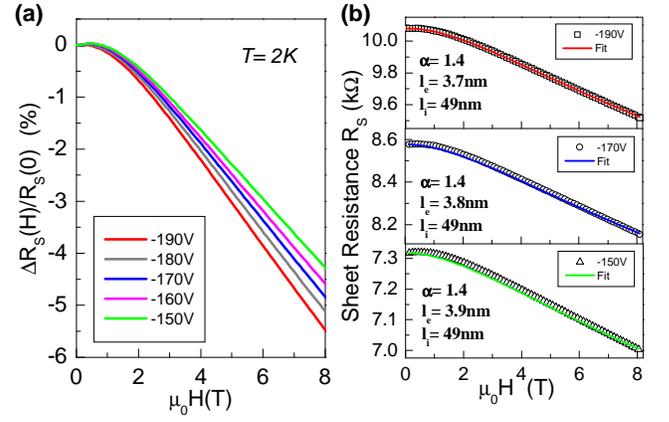}
\caption{\label{10509_Fig4} (color online) Low $\mu$ regime MR of Ar-irradiated SrTiO$_3$ at $T$=2K. (a) Evolution of MR with $V_G$. (b) $R_s(H)$ for $V_G$= -190V (squares), -170V (circles), -150V (triangles) with fits (solid lines) using Eq.5 of Ref.\cite{Bishop}. Magnitude $\alpha$, elastic $l_e$ and inelastic $l_i$ scattering lengths are indicated.}
\end{figure}

Negative MR that is consistent with the suppression of weak localization is observed for out of plane fields in the low $\mu$ regime, as shown in  Fig.\ref{10509_Fig4}(a). Weak localization arises from the constructive interference of back scattered carriers \cite{Lee_revu}. For our devices, strong confinement of the carriers near the surface can lead to such back scattering  \cite{Siemons}. $\mu_0H$ suppresses the constructive interference through additional phase shifts to the carrier wavefunctions, which is manifested as negative MR. In addition, the modulation in potential near the surface may change the Rashba spin-orbit coupling, which can also contribute to the observed crossover from positive to negative MR \cite{Koga}. In Fig.\ref{10509_Fig4}(b) we fit the $V_G$=-190V (top), -170V (middle) and -150V (bottom) MR using Eq.(5) of Ref.\cite{Bishop}. From these fits the elastic $l_e$ and inelastic $l_i$ scattering lengths have been determined, as indicated in Fig.\ref{10509_Fig4}(b), where the latter is related to the phase coherence length $L_{\phi}\sim\sqrt{l_el_i}$ \cite{Bishop}. 

Evidence for modulation of the carrier confinement can be found from the values of $l_e$ determined from the fits. As $V_G$ is changed from -190V to -150V, $\mu$ increases from 80cm$^2$/Vs to 105cm$^2$/Vs, which indicates  $\tau$=$m^*\mu e^{-1}$=$(\tau_e^{-1}$+$\tau_i^{-1})^{-1}\approx\tau_e$ increases by a factor of $\sim$1.3. An enhancement of $l_e$ would be expected since $l_e$=$v_F\tau_e$. However, $l_e$ remains virtually constant, suggesting instead that the increase in $\tau_e$ is offset by a decrease in the Fermi velocity $v_F$. We note that a weakening of the carrier confinement associated with $V_G$ would result in a decrease of the carrier density and hence $v_F$, which explains naturally the constant $l_e$ observed.  

The modulation in confinement can be understood if we consider the potential near the surface. For $V_G$=0, the carriers are confined by the built-in potential associated with oxygen vacancies. For $V_G>$0 ($V_G<$0), the electric field of the back-gate bias opposes (augments) the field of the built-in potential near the surface, resulting in weaker (stronger) confinement due to a shallower (deeper) potential. This back gate effect is further enhanced in STO due to the rapid drop in $\epsilon (E)$ with $E$ \cite{Christen}. For $V_G>$0 ($V_G<0$), the local $E$ near the surface is small (large), thus $\epsilon$ is large (small). The large $\epsilon$ essentially pulls carriers deeper into the crystal for $V_G>$0, while the reduced $\epsilon$ increases the confinement for $V_G<$0.
 
To gain insight into the effect of $\epsilon (E)$ on the confinement, we self-consistently solve the Poisson and Schr\"odinger equations for the carrier density $\rho (z)$ in the $V_G\leq$0 regime. We model our device as a 300nm thick slab where $z$=0 (-300nm) represents the surface (back side) of the device. $\rho (z)$ is obtained from $\rho (z)$=$\frac{m^*}{\pi\hbar^2}\sum_{i}(E_F-E_i)\left|\psi_i(z)\right|^2$ where the sub-band energies $E_i$ and $\psi_i(z)$ are found by numerically diagonalizing the Schr\"odinger equation. $E_F$ is determined by filling occupied sub-bands $i$ until $n_s$ is reached. The carrier potential $V$ is found by iteratively solving the Poisson $\nabla^2V$=$-\rho_{f}(z)/\epsilon (E)$ and Schr\"odinger equations. $\rho_{f}(z)$ is comprised of $\rho (z)$ as well as positive dopant charges described by $\rho_{dop}(z)\propto exp(\frac{z}{\kappa})$ for $z<$0 \cite{Yasuda}, where we have set $\kappa$=10nm. To model the effect of $V_G$ on $V$, a sheet charge density $N_G$ is placed at the back side of the slab. $N_G$ is related to $V_G$ through $N_G$=$\int^{V_G}_{0}C(V)dV$ where $C(V)$ is the measured capacitance of our device. We find that for $V_G$=-210V, $N_G\sim$2.2$\times$10$^{13}$cm$^{-2}$. The dielectric constant is calculated via $\epsilon (E(z))$= 1+($\epsilon_0^{-1})\partial P/\partial E$, where $P$ is related to $E(z)$ through the Landau-Ginzburg-Devonshire free energy \cite{Christen}. 

\begin{figure}[t]
\centering
\includegraphics[width=8.2cm] {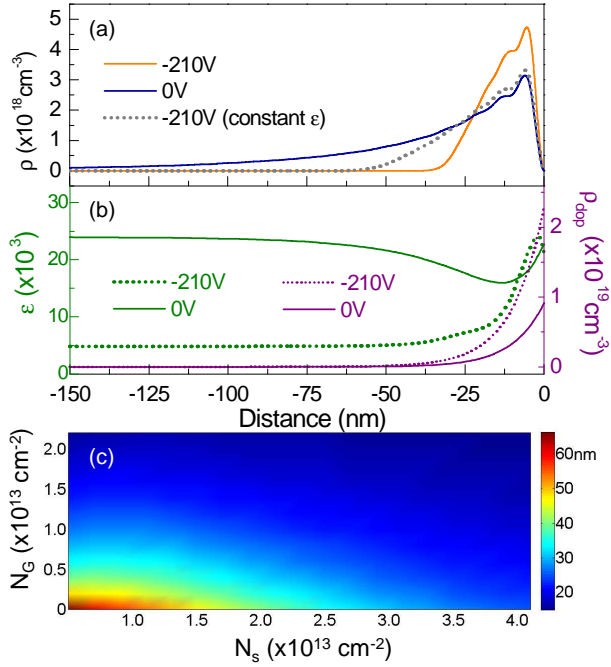}
\caption{\label{Model_Fig5} (color online) Modulation of carrier density $\rho (z)$ with $V_G$ (a) $\rho (z)$ for $V_G$= 0 (blue), -210V (orange) and -210V with $\epsilon$=24000 (grey dotted) cases. (b) $\epsilon (z)$ for $V_G$= 0 (solid green), -210V (dotted green), as well as dopant density $\rho_{dop}$ for $V_G$= 0 (solid purple), -210V (dotted purple) (c) Calculated confinement length $\zeta$ as a function of gate charge $N_G$ ($i.e.$ $V_G$) and sheet density $N_s$.}
\end{figure}
   
Figure \ref{Model_Fig5}(a) shows the calculated $\rho (z)$ for the $V_G$= 0 (blue) and -210V (orange) cases using the $n_s$ values obtained from Hall measurements. A significant difference in $\epsilon (E)$ arises between the $V_G$= 0 (green solid) and $V_G$=-210V (green dotted) cases, as shown in Fig.\ref{Model_Fig5}(b). This difference enhances the confinement, as shown in Fig.\ref{Model_Fig5}(a), where $\rho (z)$ is calculated for $V_G$=-210V using a constant $\epsilon$= 24000 (grey dashed) for comparison. We note that the confinement is also increased by the difference in $\rho_{dop}(z)$ between the $V_G$= 0 (purple solid) and $V_G$= -210V (purple dotted) cases, as shown in Fig.\ref{Model_Fig5}(b). The difference in $\rho_{dop}(z)$, required to maintain charge neutrality, can be associated with charge trapping sites that become ionized with $V_G$.   

Our simulations indicate that the modulation in confinement is also affected by $\kappa$ of $\rho_{dop}(z)$ and $n_s$. For the latter, we have solved the Poisson and Schr\"odinger equations for a range of theoretical sheet densities $N_s$. The confinement $\zeta$, quantitatively defined by $\int^{0}_{-\zeta}\rho (z)dz$=$0.7N_s$, is shown as a function of $N_G$ and $N_s$ in Fig.\ref{Model_Fig5}(c). Our simulations indicate that $\Delta\zeta$ induced by $N_G$ is greatest for small $N_s$. Regarding $\rho_{dop}(z)$, $\Delta\zeta$ decreases with smaller $\kappa$. Whereas a large $\kappa$ can generally describe $\rho_{dop}(z)$ in Ar-irradiated STO \cite{Kan,Yasuda}, a small $\kappa$ likely characterizes the LAO/STO interface \cite{Copie}. Thus, $\Delta\zeta$ with $V_G$ can be significantly smaller in the LAO/STO system, depending on the growth conditions \cite{Willmott}. The effect of $\Delta\zeta$ on superconductivity has recently been explored \cite{Hwang09}.   
   
In summary, we have induced a crossover from 3 to 2 dimensional transport in Ar-irradiated STO by applying a back gate bias. This crossover arises from an electrostatic modulation in the carrier confinement, which is enhanced by the dielectric constant of STO. Past interest has focused on utilizing the high $\epsilon (E)$ of STO to modulate $n_s$ in field effect devices \cite{McKee}. Here we have utilized $\epsilon (E)$ in an unconventional manner by combining functionalities of STO to create a mobility modulated device. With the large number of transition metal oxides available, additional devices could be developed by combining the functionalities of individual compounds.

\begin{acknowledgements}
We thank J.-M. Triscone for enlightening discussions and J. Hoffman for assistance with the figures. J.H.N acknowledges funding from NSERC. This work was supported by the National Science Foundation
under contracts MRSEC DMR-0520495 and DMR-0705799. 
\end{acknowledgements}


\begin{references}

\bibitem{Ahn03} C.H. Ahn \emph{et al.}, Nature \textbf{424}, 1015  (2003).

\bibitem{Tufte} O.N. Tufte and P.W. Chapman, Phys. Rev. \textbf{155}, 796 (1967).


\bibitem{Hwang} A. Ohtomo and H.Y. Hwang, Nature (London) \textbf{427}, 423 (2004).


\bibitem{Willmott} S.A. Pauli and P.R. Willmott, J. Phys.: Condens. Matter \textbf{20}, 264012 (2008), and references therein.

\bibitem{Hilgenkamp} A. Brinkman \emph{et al.}, Nat. Mater. \textbf{6}, 493 (2007).


\bibitem{Triscone08} A.D. Caviglia \emph{et al.}, Nature \textbf{430}, 657 (2008).

\bibitem{Triscone09} T. Schneider \emph{et al.}, cond-mat arXiv:0807.0774

\bibitem{Reagor} D.W. Reagor and V.Y. Butko, Nat. Mater.
\textbf{4}, 593 (2005).

\bibitem{Kan} D. Kan \emph{et al.}, Nat. Mater. \textbf{4}, 816 (2005). 

\bibitem{Klein} M. Schultz and L. Klein Appl. Phys. Lett. \textbf{91}, 151104 (2007).

\bibitem{Butko} V.Y. Butko \emph{et al.}, Nanotechnology \textbf{19}, 305401 (2008).

\bibitem{Siemons} W. Siemons \emph{et al.}, Phys. Rev. Lett. \textbf{98}, 196802 (2007).


\bibitem{Copie} O. Copie \emph{et al.}, Phys. Rev. Lett. \textbf{102}, 216804 (2009).

\bibitem{Fowler} T. Ando \emph{et al.}, Rev. Mod. Phys. \textbf{54}, 500 (1982).

\bibitem{Hirakawa} K. Hirakawa \emph{et al.}, Phys. Rev. Lett. \textbf{54}, 1279 (1985).



\bibitem{Christen} H.-M. Christen \emph{et al.}, Phys. Rev. B \textbf{49}, 12095 (1994).


\bibitem{Hwang09} C. Bell \emph{et al.}, cond-mat arXiv:0906.5310


\bibitem{Ionic_conductivity} An applied electric field can also cause ionic conduction of oxygen in STO at high temperatures \cite{Szot}. At $T$=2K, this scenario is highly unlikely since ionic conductivity follows $\sigma_{ion}\propto$exp($\frac{-E_a}{k_BT}$) where $E_a\sim$0.5$-$1eV \cite{Agrawal,Cordero}. 

\bibitem{Putley} E.H. Putley \emph{The Hall Effect and Semi-Conductor Physics}, Dover Publications (1968).


\bibitem{MacDonald} D.K.C. MacDonald, Nature \textbf{163}, 637 (1949); M.C. Steele, Phys. Rev. \textbf{97}, 1720 (1955); A.N. Friedman and S.H. Koenig, IBM Jour. Res. Dev. \textbf{4}, 158 (1960);


\bibitem{DepthEquation} We begin with $d$=$2R$=$2\frac{m^*v_F}{e(\mu_0H^*)}$, the classical cyclotron orbital radius $R$ at $H^*$ with Fermi velocity $v_F=\frac{\hbar}{m^*}\left(3\pi^2n\right)^{1/3}$. 
The carrier concentration $n$, is approximated to be $n=n_s/d$, which then results in Eq.\ref{Depth}.  
 
\bibitem{Lee_revu} P.A. Lee and T.V. Ramakrishnan, Rev. Mod. Phys.
\textbf{57}, 287 (1985). 

\bibitem{Koga} T. Koga \emph{et al.}, Phys. Rev. Lett. \textbf{89}, 046801 (2002).


\bibitem{Bishop} D.J. Bishop \emph{et al.} Phys. Rev. B
\textbf{26}, 773 (1982).
 

\bibitem{Yasuda} H. Yasuda \emph{et al.} Phys. Rev. B
\textbf{78}, 233202 (2008).


\bibitem{McKee} R.A. McKee \emph{et al.}, Phys. Rev. Lett. \textbf{81}, 3014 (1998).


\bibitem{Szot} K. Szot \emph{et al.}, Nat. Mater. \textbf{5}, 312 (2006).

\bibitem{Agrawal} R.C. Agrawal and R.K. Gupta, J. Mater. Sci. \textbf{34}, 1131 (1999).

\bibitem{Cordero} F. Cordero, Phys. Rev. B \textbf{76}, 172106 (2007).







\end{references}
\end{document}